\documentstyle[twoside,aps,prl,twocolumn,epsf,floats]{revtex}
\begin{document}

\title{\underline{\rm PHYSICAL REVIEW LETTERS 
~~~~~~~~~~~~~~~~~~~~~~~~~~~~~~~~~~~~~~~~~~~~~~~~~~~~~~~~~~~~~~~~~~~
\sl submitted}\\~~\\
The Extraction of Anisotropic Contributions in Turbulent Flows}
\author {Itai Arad$^1$, Brindesh Dhruva$^2$, Susan Kurien$^{1,2}$,
Victor S. L'vov$^{1,3}$, Itamar Procaccia$^1$ and K.R. Sreenivasan$^2$}
\address{$^1$Department of~~Chemical Physics, The Weizmann Institute of
Science,
Rehovot 76100, Israel\\
$^2$ Mason Laboratory and the Dept. of Physics, Yale University,
New Haven, CT 06520-8286, USA\\ $^3$ Institute of Automatization
and Electrometry Ac. Sci. of Russia, Novosibirsk 630090, Rusia}
\maketitle

\begin{abstract}
We analyze turbulent velocity signals measured by two probes in
the atmosphere, both at the height of 35 meters but displaced by 40 cm
nominally orthogonal to the mean wind. Choosing a suitable
coordinate system with respect to that of the mean wind, we derive
theoretical forms for second order structure functions, and fit them to
experimental data. We show that the effect of flow anisotropy is small
on the longitudinal component but significant on the transverse component.
The data provide an estimate of a universal exponent from among a hierarchy
that governs the decay of flow anisotropy with the scale-size.
\end{abstract}


Experimental studies of turbulent flows are usually limited in
the sense that one measures the velocity field at a single spatial
point as a function of time \cite{MY}, and uses Taylor's hypothesis
to identify velocity increments at different times with those across
spatial
length scales, $R$. The standard outputs of such measurements
are the longitudinal two-point differences of the Eulerian velocity
field and their moments, termed structure functions:
\begin{equation}
S_n(R) = \Big\langle \Big| [{\bf u}({\bf r} + {\bf R},t) - {\bf u}({\bf
r,t}) ]
\cdot {{\bbox{R}}\over R}\Big|^n \Big \rangle
\ , \label{sSn}
\end{equation}
where $\langle \cdot \rangle$ denotes averaging over time. In isotropic
homogeneous turbulence, these structure functions are observed to behave
as a power-law in $R$, $S_n(R) \sim R^{\zeta_n}$,
with apparently universal scaling exponents $\zeta_n$ \cite{Fri}.

Recent progress in measurements \cite{96CBGS} and in simulations
\cite{97Bor} begins to offer information about the tensorial nature
of structure functions. Ideally, one would like to measure the
tensorial ${\bbox{n}}$th order structure functions defined as
\begin{eqnarray}
&&S_n^{\alpha_1\dots\alpha_n}({\bbox{R}})\equiv \langle [u^{\alpha_1}
({\bbox{r}}+{\bbox{R}})-u^{\alpha_1}({\bbox{r}})]
\nonumber\\&\times&[u^{\alpha_2}({\bbox{r}}+{\bbox{R}})-u^{\alpha_2}
({\bbox{r}})]\dots[u^{\alpha_n}({\bbox{r}}+{\bbox{R}})-
u^{\alpha_n}({\bbox{r}})]\rangle \ , \label{Sn}
\end{eqnarray}
where the superscript $\alpha_i$ indicates the velocity component
in the direction $i$. Such information should be useful in studying
the anisotropic effects induced by all practical means of forcing.
In analyzing
experimental data the model of ``homogeneous, isotropic small-scale" is
universally adopted, but it is important to examine the relevance
of this model
for realistic flows. One of the points of this Letter is that keeping
the tensorial information helps significantly in disentangling different
scaling contributions to structure functions \cite{97LPP}. Especially
when anisotropy might lead to different scaling exponents for different
tensorial components, a careful study of the various contributions is
needed. We
will show below that atmospheric measurements contain important
anisotropic contributions to one type of transverse structure functions.

In this Letter we analyze measurements in atmospheric turbulence
at Taylor microscale Reynolds numbers of about 20,000 \cite{sreeni}.
The data were acquired
simultaneously from two single-wire probes separated by 40 cm
nominally orthogonal to the
mean wind direction. The two probes were mounted at a height of about
35 m above the ground on a meteorological tower at the
Brookhaven National Laboratory. The hot-wires, about 0.7 mm in
length and 6 $\mu$m in diameter, were calibrated just prior
to mounting them on the tower (and checked immediately after
dismounting), and operated on DISA 55M01 constant-temperature
anemometers. The frequency response of the hot-wires was typically
good up to 20 kHz. The voltages from the anemometers were suitably
low-pass
filtered and digitized. The voltages were constantly monitored
on an oscilloscope to ensure that they did not exceed the digitizer
limits. Also monitored on-line were spectra from an HP 3561A Dynamic
Signal Analyzer. The wind speed and direction were independently
monitored by a vane anemometer mounted a few meters away from the
tower. The real-time duration
of data record was limited
by the degree of constancy of the speed and direction of the wind.
The Kolmogorov scale was about 0.45. Table~1 lists a few relevant facts the 
data records listed here (this being part of a much larger set). 
The various symbols have the following
 meanings: $\overline U$ = local mean velocity, $u^{\prime}$ =
root-mean-square velocity, $\langle \varepsilon \rangle$ = energy
dissipation obtained by the assumption of local isotropy and Taylor's
hypothesis, $\eta$ and $\lambda$ are the Kolmogorov and Taylor length
scales, respectively, the micro-scale Reynolds number $R_{\lambda}
\equiv u^{\prime} \lambda/\nu$, and $f_s$ is the sampling frequency.
\begin{table}
\begin{tabular} {|c|c|c|c|c|c|c|c|}
$\overline U$ & $u^\prime$ &$10^2 \langle \varepsilon \rangle $,        & $\eta$ &
$\lambda$ & $R_{\lambda}$ & $f_s,$ per & \# of \\ ms$^{-1}$ & ms
$^{-1}$ & m $^2$ s$^{-3}$ & mm & cm & & channel, Hz & samples\\ \hline
8.3 & 2.30 & $7.8 $ & 0.45 & 13 & 19,500 & 5,000 & $ 4 \times 10^7$\\
\end{tabular}
\caption{ Basic information about the data analyzed in this paper.}
\end{table}

As a first step we tested whether the separation between the two probes
was indeed orthogonal to the mean wind. To this end we computed
the cross-correlation function $\langle u_1(t+\tau)u_2(t)\rangle$ where
subscripts 1 and 2 refer to the two probes. If the separation were precisely
orthogonal to the mean wind, this quantity should be maximum for $\tau=0$.
Instead, we found the maximum at 0.03 sec, implying
that the separation was not precisely orthogonal to the mean wind. To
correct for this effect, the data from the second probe were time-shifted
by 0.03
seconds. This amounts to a change in the actual value of the orthogonal
distance, and we computed this effective distance to be $\Delta
\approx 31$ cm. Next we tested the isotropy of the flow for separations
of the order of $\Delta$. Define the ``transverse" structure function
across $\Delta$ as $S_T(\Delta)\equiv\langle [u_1(\bar U t)-u_2(\bar U
t)]^2\rangle$
and the ``longitudinal" structure function as $S_L(\Delta)\equiv \langle
[u_1(\bar U t+\bar
U t_\Delta)-u_1(\bar U t)]^2\rangle$ where $t_\Delta=\Delta/\bar U$. If the
flow were isotropic we would expect \cite{MY} \begin{equation}
S_T(\Delta)=S_L(\Delta)+{\Delta\over 2}{\partial S_L(\Delta)\over
\partial \Delta} \ . \label{Stl} \end{equation}
In the isotropic state both components scale with the same exponent,
$S_{T,L} (\Delta)\propto \Delta^{\zeta_2}$, and their ratio is computed
from (\ref{Stl}) to be $1+\zeta_2/2\approx 1.35$ where $\zeta_2\approx
0.69$ (see below). The experimental ratio was found to
be 1.86, indicating some 40\% anisotropy at this scale \cite{fn1}; we
expect even more anisotropy on larger scales.

To obtain a theoretical form of the structure function tensor we first
select a natural coordinate system. An obvious choice is the mean-wind
direction $n$ along the $\alpha=3$ axis. A second axis is given by the
separation vector ${\bbox{\Delta}}$ between the two probes. It turns out that
{\em for this particular geometrical configuration} such a two-dimensional
coordinate system is sufficient to describe all the non-zero tensor components.
This is to be expected since all measurements of velocity and
all separations are contained in the plane (${\bbox{n}}$, ${\bbox{\Delta}}$). We can
envision
a more general experimental setup which measures velocity components
perpendicular to the ground or in which the probes are separated in the
vertical direction. In such a situation, we would need to take into
consideration tensor components arising from the existence of a non-trivial
third direction orthogonal to the (${\bbox{n}}$, ${\bbox{\Delta}}$) plane.
We expect
that such experiments will in the future allow us to study the
antisymmetric components of the tensor (see below).

To continue, we need to write down the tensor form for the general
second order structure function (defined by Eq.~(\ref{Sn}) for $n=2$)
in terms of irreducible representations of the SO(3) rotation group.
This tensor can be written in terms of the representations of the direct
product of two three-dimensional Euclidean vector spaces (for the
indices $\alpha$, $\beta$) and the space of continuous functions on
the unit sphere (for the direction of ${\bbox{R}}$) \cite{98ALP}. The latter is
spanned by the spherical harmonics $Y_{l,m}$, and the representations
of the product space are indexed by $j$, denoting a $2j+1$ dimensional
irreducible representation. Every such representation is associated
with a scalar function $c_j(R)$ which is expected to scale with a
universal exponent $\zeta_2^{(j)}$; the latter is an increasing function of
$j$ and
$\zeta_2^{(0)}=\zeta_2$. Previous theoretical considerations \cite{73Les}
led to the
estimates $\zeta_2^{(1)}\approx 1,~\zeta_2^{(2)}\approx 4/3$. We are
interested in
relatively modest anisotropies, and so focus on the lowest order correction to
the isotropic ($j=0$) contribution. In other words, we write
\begin{equation}
S^{\alpha\beta}({\bf R}) = S_{j=0}^{\alpha\beta}({\bf R}) +S_{j=2}^
{\alpha\beta}({\bf R})+\dots
\end{equation}
We do not have a $j=1$ term since the possible contributions to it vanish 
either because of parity considerations (the structure function itself is 
even in R) or by the incompressibility constraint. Now the most general 
form of the tensor can be written down by inspection. The case $j=0$ is 
well known, and we write it as
\begin{equation}
S_{j=0}^{\alpha\beta}({\bf R})=c_0(R) \left[(2+\zeta_2) \delta^
{\alpha\beta}-\zeta_2{R^\alpha
R^\beta\over R^2}\right]\ , \label{Siso} \end{equation}
where $c_0(R)=c_0 R^{\zeta_2}$, and $c_0$ is a non-universal numerical
constant that needs to be fitted to the data. The $j=2$ component can be
written as
\FL
\begin{eqnarray}\label{j2tens}
&S& ^{\alpha\beta}_{j=2} ({\bf R}) = c_2(R)\Big [a\delta^{\alpha\beta}
+ b{R^\alpha R^\beta \over R^2}
+d{\delta^{\alpha\beta}({\bbox{n}}\cdot {\bbox{R}})^2\over R^2} \\ 
&+&e{R^\alpha R^\beta
({\bbox{n}}\cdot {\bbox{R}})^2\over R^4}
+f n^\alpha n^\beta+{g({\bbox{n}}\cdot {\bbox{R}}) \over 2R^2} 
(R^\alpha n^\beta + R^\beta n^\alpha)
\Big ]
\nonumber
\end{eqnarray}
where $c_2(R)=R^{\zeta_2^{(2)}}$. Here $a,~b,~d,~e,~f$ and $g$ are unknown
coefficients which are independent of $R$. This form can be further
reduced by imposing the conditions of incompressibility and
orthogonality with the $j=0$ part of the tensor. This leaves us
with only two independent coefficients of the form \FL
\begin{eqnarray}
&S&^{\alpha\beta}_{j=2} ({\bf R}) =
aR^{\zeta_2^{(2)}}\Big[(\zeta_2^{(2)}
-2)\delta^{\alpha\beta} -
\zeta_2^{(2)}(\zeta_2^{(2)}+6)\nonumber\\
&\times&\delta^{\alpha\beta}{({\bbox{n}}\cdot {\bbox{R}})^2
\over R^2}+2\zeta_2^{(2)}(\zeta_2^{(2)}-2){R^\alpha R^\beta({\bbox{n}}\cdot 
{\bbox{R}})^2\over R^4}\nonumber\\ 
&+&([\zeta_2^{(2)}]^2+3\zeta_2^{(2)}+6)n^\alpha n^\beta \nonumber\\
&-&{\zeta_2^{(2)}(\zeta_2^{(2)}-2)\over R^2}(R^\alpha n^\beta +
R^\beta n^\alpha)({\bbox{n}}\cdot {\bbox{R}})\Big]\label{finalform}\\ 
&+& bR^{\zeta_2^{(2)}}\Big
[-(\zeta_2^{(2)} +3)(\zeta_2^{(2)}+2)\delta^
{\alpha\beta}({\bbox{n}}\cdot {\bbox{R}})^2 +
{R^\alpha R^\beta \over R^2} \nonumber\\
&+& (\zeta_2^{(2)} +3)
(\zeta_2^{(2)}+2)n^\alpha n^\beta + (2\zeta_2^ {(2)}+1)(\zeta_2^{(2)}-2)
\nonumber\\
&\times&{{R^\alpha}{R^\beta}{({\bbox{n}}\cdot {\bbox{R}})^2} \over R^4}-
([\zeta_2^{(2)}]^2- 4)(R^\alpha n^\beta + R^\beta n^\alpha)
({\bbox{n}}\cdot {\bbox{R}})\Big] \ .
\nonumber
\end{eqnarray}
Finally, we note that in the present experimental set-up only the component
of the velocity in the direction of ${\bbox{n}}$ is measured. In the coordinate
system chosen above we can read from (\ref{finalform})
the relevant component as
\FL
\begin{eqnarray}\label{ourcomp}
S^{33}(R,\theta)&=&S^{33}_{j=0}(R,\theta)+ S^{33}_{j=2}(R,\theta)\\
=c_0\left({R\over \Delta}\right)^{\zeta_2}& \Big[& 2+\zeta_2-\zeta_2
\cos^2\theta\Big]\nonumber\\
+a\left({R\over \Delta}\right)^{\zeta_2^{(2)}}& \Big[& (\zeta_2^{(2)}+2)^2
-\zeta_2^{(2)}
(3\zeta_2^{(2)}+2)\cos^2\theta\nonumber\\ &+&2\zeta_2^{(2)}(\zeta_2^{(2)}
-2)\cos^4\theta\Big] \nonumber\\+b\left({R\over
\Delta}\right)^{\zeta_2^{(2)}}&\Big[&(\zeta_2^{(2)}+2) (\zeta_2^{(2)}+3)-
\zeta_2^{(2)}(3\zeta_2^{(2)}+4)\cos^2\theta\nonumber\\&+&(2\zeta_2^{( 2)}+1)
(\zeta_2^{(2)}-2)\cos^4\theta\Big] \ . \nonumber \end{eqnarray}
Here $\theta$ is the angle between ${\bbox{R}}$ and ${\bbox{n}}$, and we
normalized $R$ by $\Delta$ making all our coefficients dimensional,
with units of (m/sec)$^2$.

To fit our experimental results we converted, using Taylor's
hypothesis, the structure functions computed from time differences for
a single probe, and cross differences between the two probes, to
components of the form (\ref{ourcomp}) with $\theta=0$ and with
variable $\theta$, respectively. In other words,
\begin{equation}
S^{33}(R,\theta=0)=\langle [u_1
(\bar U t+\bar Ut_R)-u_1(\bar Ut)]^2\rangle\ ,
\end{equation}
where $t_R\equiv R/\bar U$, and
\begin{equation}
S^{33}(R,\theta)=\langle [u_1(\bar U t
+\bar Ut_{\tilde R})-u_2(\bar Ut)]^2\rangle \end{equation}
Here $\theta=\arctan(\Delta/\bar Ut_{\tilde R})$, $t_{\tilde R}=\tilde
R/\bar U$, and $R=\sqrt{\Delta^2+(\bar Ut_{\tilde R})^2}$. We then found the
three unknown coefficients $c_0,~a$
and $b$, and the two exponents $\zeta_2$ and $\zeta_2^{(2)}$ by a nonlinear
fit to the form (\ref{ourcomp}). The fitted numbers are presented in Table~2.
\begin{table}
\begin{tabular} {|c|c|c|c|c|}
$\zeta_2$&$\zeta_2^{(2)}$&$c_0$&$a$&$b$\\ \hline 0.69&1.36&0.112&-0.052&0.050\\
\end{tabular}
\caption{The scaling exponents and the three coefficients in units of
(m/sec)$^2$ as determined from the nonlinear fit of Eqs. (9) to the data.}
\end{table}

\begin{figure}
\epsfxsize=8.5truecm
\epsfbox{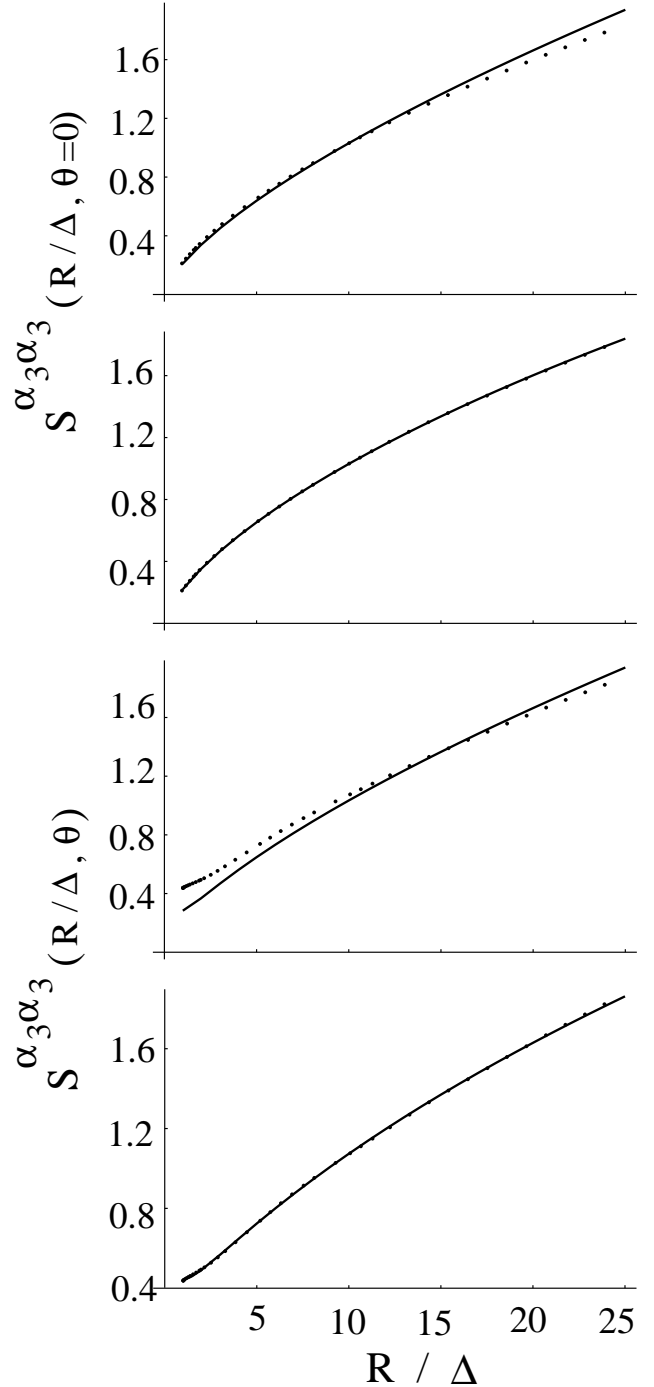}
\caption{The structure functions $S^{33}$ for $\theta=0$ in panels
(a) and (b), and for non-zero $\theta$ in panels (c) and (d). The dots are
for experimental data and the line is the analytic fit. Panels (a) and (c)
present fits to the $j=0$ component only, and panels (b) and (d) to
components $j=0$ and $j=2$ together.} \label{Fig.1}
\end{figure}

\begin{figure}
\epsfxsize=9.0truecm
\epsfbox{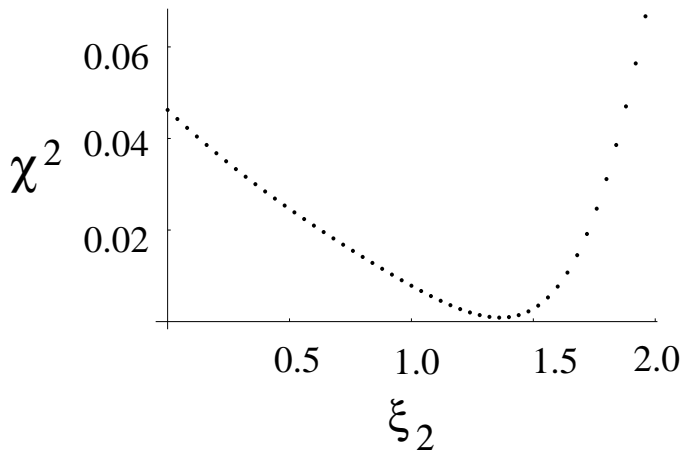}
\caption{The determination of the exponent $\zeta_2^{(2)}$ from a
least-square fit
of $S^{33}(R,\theta)$ to its analytic form. The numerical value of the best
fit, $\xi=1.36\pm 0.1$ is in close agreement with the theoretical
expectation of 4/3 (without intermittency corrections).}
\label{Fig.2}
\end{figure}

One surprise of the analysis is that the $\theta=0$ (purely longitudinal)
structure function is hardly affected by the anisotropy.
The reason is the very close numerical absolute values of the coefficients
$a$ and $b$. In the case $\theta=0$ the two tensor forms multiplied by $a$ and
$b$ coincide, and the $j=2$ contribution becomes
very small. We believe that this is the main reason that the effect of
anisotropy
has been disregarded by and large in
atmospheric experiments. The component
$S^{33}(R,\theta=0)$ can be reasonably fitted by a power law with single
exponent.
In fact, the value of $\zeta_2=0.69$ quoted above can be obtained from such
a simple
fit, and as long as one measures only this component, there is no reason for a
more sophisticated analysis. In fact, $\zeta_2$is so well estimated that we
can safely take it as fixed when we perform the simultaneous fit of all the
other unknowns.

On the other hand, the finite $\theta$ components are strongly affected
by the anisotropy, and the inclusion of the second exponent $\zeta_2^{(2)}$
is essential for a good fit. In Fig.~1 we exhibit various
quantities as a function of $R/\Delta$; the first panel shows $S^{33}(R,
\theta=0)$
with its best fit to the $j=0$ contributions, and in panel (b) with its
best fit
to the sum of $j=0$ and $j=2$. The differences, though small as noted
previously,
are enough to produce an improved agreement. Similarly in panels (c) and
(d) we show
$S^{33}(R,\theta)$ with its best fit to the $j=0$ and the sum of $j=0$ and
$j=2$,
respectively. It is clear that panel (c) does not give an adequate fit in the
``inertial range", as estimated from panel (a) to range up to, say, 15$\Delta$.
If we tried to read a log-log slope from the data in panel (c) we would get an
apparent scaling exponent $\zeta_2$ considerably smaller than 0.69. The
excellent
fits in panels (b) and (d) are a good support to the present mode of analysis.

To our knowledge, this determination of $\zeta_2^{(2)}$ is the first instance
of finding an exponent describing the degree of anisotropy. The close
agreement with the theoretical expectation of 4/3 (e.g., Ref.~\cite{lumley})
is a strong
indication that this exponent is universal. In Fig.~2 we present $\chi^2$
(the sum of the squares of the differences between the experimental data
and the fitted values) as a function of $\zeta_2^{(2)}$.
The optimal value of this exponent and the uncertainty determined from
this plot is $\zeta_2^{(2)}\approx 1.36\pm 0.1$.

It should be understood that the exponent $\zeta_2^{(2)}$ (and also
$\zeta_2^{(1)}$ that is
unavailable from the present measurements) are just the smallest exponents
in the hierarchy $\zeta_2^{(j)}$ that characterizes higher order irreducible
representations indexed by $j$. The study of these exponents is in
its infancy, and considerable experimental and theoretical effort is needed
to reach firm conclusions regarding their universality and numerical values.
We expect the exponents to be a non-decreasing function of $j$, explaining
why the highest values of $j$ are being peeled off quickly when $R$ decreases.
Nevertheless, the lower order values of $\zeta_2^{(j)}$ can be measured and
computed. It is the program of the present authors to proceed in this direction.

At Weizmann, the work was supported by the Basic Research Fund administered
by the Israeli Academy of Sciences, the US-Israel Binational Science
Foundation and the Naftali and Anna Backenroth-Bronicki Fund for Research
in Chaos and Complexity. At Yale, it was supported by the National Science
Foundation grant DMR-95-29609, and the Yale-Weizmann Exchange Program.

\end{document}